\documentclass[superscriptaddress,showpacs,twocolumn,prl]{revtex4}
\usepackage{graphicx,amsfonts}
\usepackage{epsfig,amsmath}

\begin{document}

\newcommand{\atanh}
{\operatorname{atanh}}
\newcommand{\ArcTan}
{\operatorname{ArcTan}}
\newcommand{\ArcCoth}
{\operatorname{ArcCoth}}
\newcommand{\Erf}
{\operatorname{Erf}}
\newcommand{\Erfi}
{\operatorname{Erfi}}
\newcommand{\Ei}
{\operatorname{Ei}}

\title{Aging in the glass phase of a 2D random
periodic elastic system}
\author{Gregory Schehr}
\affiliation{Theoretische Physik Universit\"at des Saarlandes
66041 Saarbr\"ucken Germany}
\author{Pierre Le Doussal}
\affiliation{CNRS-Laboratoire de Physique
Th\'eorique de l'Ecole Normale Sup\'erieure, 24 Rue Lhomond 75231
Paris, France}


\date{\today}
\begin{abstract}
Using RG we investigate the non-equilibrium relaxation of
the (Cardy-Ostlund) 2D random Sine-Gordon model,
which describes pinned arrays of lines. Its statics exhibits a
marginal ($\theta=0$) glass phase for $T<T_g$
described by a line of fixed points. We obtain the universal
scaling functions for two-time dynamical response and correlations
near $T_g$ for various initial conditions,
as well as the autocorrelation exponent. The
fluctuation dissipation ratio is found to be non-trivial
and continuously dependent on $T$.
\end{abstract}
\maketitle

Tremendous progress has been achieved in recent years
in the detailed understanding of the off equilibrium
relaxational dynamics (e.g. coarsening) of pure systems.
For instance the scaling forms of response and
correlations are well characterized and
several autocorrelation and persistence exponents
have been computed using powerful methods
\cite{revuecoars}.
By contrast, analytical
studies of aging properties of random and complex
systems have been mostly achieved within mean field
theory \cite{leto_leshouches}. These works have unveiled the existence of
non trivial fluctuation dissipation ratios (FDR), $X$. These generalize  
to non-equilibrium situations the fluctuation
dissipation relation between 
(disorder averaged) integrated response and correlation
\cite{leto_leshouches} and can be interpreted as an effective
temperature $T_{\text{eff}} = T/X$ \cite{leticia_teff}. 
These were later
investigated beyond mean field for pure systems.
Although they appear to be trivial for some pure domain growth
processes \cite{domaingrowth}, they were later found to be novel
universal quantities at pure critical points \cite{criticalfdr, calabrese_on}.
 
The description of glassy states beyond mean field
needs to take into account very broad distribution of
relaxation times originating from rare sample to sample
fluctuations and distribution of barrier heights \cite{balentspld}.
In this context it is not obvious how (or whether)
the off equilibrium dynamical scaling valid for pure systems
extends to this case. Also it is not obvious that
various definitions of FDR (using different observables, or different
configurational averages) are equivalent and meaningful
\cite{chamon,rfim,mayer_fdr_ising}. Some results were
obtained for the random mass Ising critical point,
via RG. The dynamical ($z$) and autocorrelation ($\theta_C$) exponent
were computed \cite{randommass1}. A
definition of the FDR (for the zero mode) was
found to be non trivial \cite{randommass2} and obtained to one loop.

Whether such results hold {\it within a glass phase}, away from
a phase transition, is still an outstanding question. A prototype
system which exhibits a glass phase, while analytically more tractable,
is the disordered elastic manifold in a random potential
\cite{revuevortex}. It  
describes a variety of physical situations, both in or
out of equilibrium, ranging
from visualization of pinned domain wall relaxation
in magnets \cite{ferre,krusin}, 
non equilibrium transport in electronic glasses \cite{coulombglass}
to mesoscopic fluctuations in vortex physics \cite{bolle}. 
Here we investigate the case of a periodic
manifold, a case which can be mapped \cite{natter1,brG}
onto the so-called Cardy-Ostlund (CO), 2D- random Sine Gordon
model, \cite{co} defined by the hamiltonian:
\begin{equation}
H[\phi] = \int d^2 x ( \frac{1}{2} (\nabla_x \phi(x))^2 - \text{Re}(
\xi(x) e^{i \phi(x)} )    \label{HamCO} 
\end{equation}
where $\phi(x) \in ]-\infty, + \infty[$ is an XY phase (excluded
vortices), $\xi(x)$  
is a quenched gaussian random field, i.e. a complex variable, with random phase
and $\overline{\xi(x) \xi(x')^*}= g \delta^{(2)}(x-x')$. The statics
    of (\ref{HamCO})  
has been extensively studied analytically and numerically, in the context of
planar flux line arrays (with displacement field $u=a \phi/(2 \pi)$ and
mean spacing $a$) and solid on solid models with disordered substrates. 
It is known to exhibit a glass phase \cite{toner,fisherhwa} 
below $T_g$ ($=4 \pi$ ) described by a line 
of fixed points perturbatively controlled by the small parameter $\tau =
(T_g-T)/T_g$ (see however \cite{sanchez_rough}). By contrast, its (non
driven) non equilibrium dynamics has only been studied within gaussian
variational approximation GVA \cite{cule_mf}, known already in statics
to yield unreliable results for some observables
\cite{footnotenew}. There is thus
the need for a controlled study via RG.

In this paper we study the relaxation dynamics for the CO model in the
glass phase $T<T_g$ 
starting from various initial conditions at $t_i=0$
(e.g. with the same correlations as the pure model
at equilibrium at temperature $T'$). We compute, to lowest order
in $\tau$ using RG along the fixed line, the
two time ($t'<t$) response ${\cal R}$, as well as the 
connected $\tilde {\cal C}$ and, respectively, 
off-diagonal ${\cal C}$ correlations. These are
found to be characterized by ($T'$-dependent) universal 
scaling functions of $q^z (t-t')$ and $t/t'$,
$q$ being the wavevector. We
find that an  
autocorrelation exponent can be consistently defined, 
i.e. with $\theta_{\tilde C} = \theta_{R}$,
only from the connected correlation (while the decay of ${\cal C}$ is
too slow). 
Similarly, the associated FDR, found 
to be a scaling function, reaches at large time separation 
($t/t' \to \infty$) a non trivial (and to one loop, $q$-independent)
limit $X_\infty$ which is non zero only for the
connected correlation. 
This sheds light on relevant definition of the FDR in a glass phase.

Let us start by a short discussion of the statics. The 
$2$-point correlation function exhibits anomalous
growth 
\begin{eqnarray}\label{log_carre}
\overline{\langle (\phi(x) - \phi(0))^2 \rangle} = A(\tau) \ln^2 (x) +
	 {\cal O}(\ln x)
\end{eqnarray} 
The RG to one loop \cite{dc} yields the universal \cite{footnote1}
$A(\tau) =2 \tau^2 + O(\tau^3)$. Numerical simulations near $T_c$
\cite{simu_tc} and at $T=0$ agree qualitatively,
and yield $A(\tau=1)= a_2 = 0.57$ \cite{rieger}
and $A(\tau=1)= 2 (2 \pi)^2 B  = 0.51$ \cite{middleton}.
A recent work \cite{ludwig} claims an exact solution for
the $2$-point correlation using a conjectured mapping 
onto disordered free fermion model. The result, translated in the
present variables 
would yield $A(\tau)=2 \tau^2 (1-\tau^2)$. This however
is clearly incompatible with numerics, showing that there
is more (non perturbative ?) physics to be understood 
even in the statics of this model. A
more direct  
signature of the glassy nature of the phase is the 
sample to sample susceptibility fluctuations \cite{fisherhwa} 
which are found (within one loop RG) to be universal
and $\sim {\cal O}(\tau)$ along the fixed line. The free 
energy exponent in this glass phase is $\theta=0$,
indicating a free energy landscape with logarithmic
roughness. This is consistent with the findings 
in the equilibrium dynamics: an anomalous diffusion exponent \cite{gold},
continuously varying along the fixed line 
has been computed in RG which indicates a logarithmic growth of energy
barriers with scale. These
properties are 
caracteristic of a marginal glass (by contrast with 
the case $\theta>0$ described by a $T=0$
fixed point). It is reminiscent of a related and
simpler case of a vortex in a random gauge XY model
where a freezing transition (at $z_c=4$) was found along
a line of fixed points \cite{marginal_freezing}. 

The relaxational dynamics of the CO model (\ref{HamCO}) is 
described by a Langevin equation
\begin{equation}
\eta \frac{\partial}{\partial t} \phi(x,t) = - \frac{\delta
  H[\phi(x,t)]}{\delta 
\phi(x,t)} + \zeta(x,t) \label{Eq_Langevin}
\end{equation}
where the thermal noise $\zeta(x,t)$ is such that $\langle
\zeta(x,t)\rangle = 0 $, 
$\langle \zeta(x,t) \zeta(x',t') \rangle = 2\eta T \delta^{(2)}(x-x')
\delta(t-t')$ where $T < T_g$ is the temperature and $\eta$
the friction ($\eta = 1$ in the following). The system at
initial time $t_i$ ($ = 0$) 
is prepared in an equilibrium state of (\ref{HamCO})
{\it without} disorder at temperature $T' = \epsilon T$,
$[{\phi}_{q,t=0} {\phi}_{-q,t=0}]_i = T'q^{-2}$, 
${\phi}_{q,t} $ being the Fourier transform, w.r.t. space
coordinates, of the field
$\phi(x,t)$. Since the disorder is irrelevant above
$T_g$, this choice of initial condition for $T' > T_g$
describes a quench from a high temperature phase to
the glass phase ($T< T_g$). A quench to high temperature
($T>T_g$) is studied in \cite{berthier_xy}. 
   
We will focus on the correlation ${\cal
C}^q_{tt'}$ and the connected (w.r.t. the thermal fluctuations) 
correlation ${\tilde{\cal C}}^q_{tt'}$  
\begin{eqnarray}\label{def_C}
{\cal{C}}^q_{tt'} &=&[\overline{\langle {\phi}_{qt}
      {\phi}_{-qt'}  \rangle}]_i \nonumber \\
{\tilde{\cal C}}^q_{tt'} &=& [\overline{\langle {\phi}_{qt}
      {\phi}_{-qt'}  \rangle}]_i - 
[\overline{{\langle {\phi}_{qt}\rangle} {\langle
      {\phi}_{-qt'}\rangle}}]_i  
\end{eqnarray}
and the response ${\cal R}^q_{tt'}$ to a small external
field ${f}_{-qt'}$  
\begin{eqnarray}\label{def_R}
{\cal R}^q_{tt'} = \overline{\left[\frac{\delta \langle
  {\phi}_{qt} \rangle}{\delta 
  {f}_{-qt'}}\right]_i } \quad, \quad t > t'
\end{eqnarray}
where $\overline{..}$, $\langle .. \rangle $ and
$[..]_i$ denote respectively averages w.r.t. disorder,
thermal fluctuations and initial conditions. We focus on the 
FDR ${\cal X}^q_{tt'}$  associated to 
the observable $\phi$ \cite{leto_leshouches}:
\begin{equation}
\frac{1}{{\cal X}^q_{tt'}} = \frac{\partial_{t'} \tilde{\cal
    C}^q_{tt'}}{T {\cal R}^q_{tt'}}
 \label{def_FDR}
\end{equation}
defined such that ${{\cal X}^q_{tt'}} = 1$ at
equilibrium, {\it i.e.} when response and correlations depend only on
$t-t'$ \cite{foot_X}.

The dynamics (\ref{Eq_Langevin}) of the CO model (\ref{HamCO}) is then
studied using the standard MSR formalism
\cite{martin_siggia_rose,janssen_noneq_rg}, 
using the Ito prescription. The correlations (\ref{def_C}) and response
(\ref{def_R}) are then obtained from a dynamical (disorder averaged)
generating functional or, equivalently, as
functional derivatives of the corresponding dynamical {\it effective}
action $\Gamma$. This functional can be perturbatively computed
\cite{schehr_co_pre} 
using the Exact RG equation associated to the
multi-local operators expansion introduced in \cite{chauve}.
It allows to handle arbitrary cutoff functions $c(q^2/2 \Lambda_0^2)$ and check
universality, independence w.r.t. $c(x)$ and the ultraviolet scale
$\Lambda_0$.  
It describes the evolution of $\Gamma$
when an additional infrared cut-off $\Lambda_l$ is lowered from 
$\Lambda_0$ to its 
final value $\Lambda_l \to 0$ where a fixed point of order
${\cal O}(\tau)$ is reached. In this
limit, one obtains ${\cal R}^q_{tt'}$ and $\tilde{\cal C}^q_{tt'}$
(for $t>t'$) from
\begin{equation}
\partial_{t} {{\cal R}}_{tt'}^{q} + (q^2 - \int_{t_i}^t d {t_1} \Sigma_{tt_1})
{{\cal R}}_{tt'}^{q} +
\int_{t_i}^t d {t_1} \Sigma_{tt_1}
{{\cal R}}_{t_1t'}^{q} =
\delta(t-t')   \label{Eq_R}
\end{equation}
\begin{equation}
\tilde{\cal C}^{ {q}}_{ {t} {t'}} =
2T\int_{t_i}^{{t'}} dt_1 {{\cal R}}^{ {q}}_{ {t}t_1}{{\cal
R}}^{ {q}}_{ {t'}t_1} 
+ \int_{t_i}^{ {t}} dt_1 \int_{t_i}^{ {t'}} dt_2{{\cal
R}}^{ {q}}_{ {t} t_1} D^c_{t_1t_2}
{{\cal R}}^{ {q}}_{ {t'} t_2}   
\label{expr_cc_dis_ci}
\end{equation}
where the self energy $\Sigma_{t_1t_2}$ and the noise-disorder kernel 
$D^c_{t_1 t_2}$ are directly obtained
from $\Gamma$ at the fixed point \cite{foot3}. One finds (for details
see \cite{schehr_co_pre,
schehr_new}):
\begin{eqnarray}
&& \Sigma_{t t'} = - 2^{\epsilon-1} e^{\gamma_E} \tau R_{t t'}^{x=0} 
e^{- \frac{1}{2} ( B^{(0)}_{t t'} + B^{(d)}_{t t'} )}  \\
&& D^c_{t t'} = T_g 2^{\epsilon-1} e^{\gamma_E} \tau
e^{- \frac{1}{2} ( B^{(0)}_{t t'} + B^{(d)}_{t t'} )} (1-e^{- C^{(d)}_{t t'}})
\end{eqnarray} 
where $\gamma_E$ is the Euler constant,
$B^{(d)}_{t t'} = C^{(d)}_{t t} +  C^{(d)}_{t' t'} - 2  C^{(d)}_{t t'}$,
$C^{(d)}_{t t'} = \gamma(t+t') - \gamma(|t-t'|)$ is the bare Dirichlet
propagator at coinciding points, 
$B^{(0)}_{tt'} = \epsilon (2\gamma(t+t')-\gamma(t)-\gamma(t'))$, arising
from the average over the initial condition.
We have defined $\gamma(t)=
\frac{T}{4 \pi} \int_a \hat c(a) \ln(\Lambda_0^2 t +\frac{a}{2})$,
using the parameterization $c(x)=\int_a \hat c(a) e^{-a x}$
for the cutoff function, and denote 
$R_{t t'}^{x=0} = \theta(t-t') \int_a \hat c(a)  (\Lambda_0^2 (t-t')+\frac{a}{2})^{-1}$
the bare response at coinciding points. Up to a boundary term,
the correlation $C^q_{tt'}$ satisfies (\ref{expr_cc_dis_ci}) 
setting the $e^{- C^{(d)}_{t t'}}$ term to zero in the
above expression for $D^c_{tt'}$. 

Studying Eq. (\ref{Eq_R}) (setting $t_i = 0$), 
in the scaling regime $q/\Lambda_0 \ll
1$, $\Lambda_0^2 t, \Lambda_0^2 t' \gg 1$ and keeping:
\begin{equation}
v \sim {q^z}(t-t') \quad , \quad u = \frac{t}{t'} 
\end{equation}
fixed, one finds a solution consistent with the
scaling~form:
\begin{equation}
{\cal R}^{{q}}_{{t}{t'}} =
{q}^{z-2}\left( \frac{{t}}{{t'}} \right)^{\theta_R}
F_{R}({q}^z({t}-{t'}),{t}/{t'})
\label{scaling_resp}
\end{equation}
similar to the form obtained
for the response for critical systems \cite{janssen_noneq_rg,
randommass1,picone_autoresponse}.  
Here the two exponents $z$ and $\theta_R$ are identified
from the logarithmic singularities of the scaling function,
respectively at $u \to 1$ and $u \to \infty$. We find
that $z$ identifies with the equilibrium dynamical
exponent $z-2 = 2 e^{\gamma_E} \tau + O( \tau^2)$
and is thus, as expected independent of initial
conditions under study. The exponent $\theta_R$ in
(\ref{scaling_resp}), 
characteristic of long time non-equilibrium relaxation 
is also independent of $\epsilon$:
\begin{equation}\label{theta_R}
\theta_R = e^{\gamma_E} \tau + {\cal O} (\tau^2)
\end{equation}
One also finds, with this choice of $\theta_R$,
that $\lim_{u \to  \infty} F_R(v,u) = F_{R\infty}(v)$.
The full scaling function $F_R(v,u)$ however 
depends on the initial conditions and is universal (up to a single
overall non-universal length scale $q \to \lambda q$).
Its explicit expression, given in \cite{schehr_new},
contains both non-equilibrium and equilibrium
($u \to 1$) regimes. We have checked to this order
that it naturally
splits into $F_{R}(v,u) = F_R^{\text{eq}}(v) 
+ F^{\text{noneq}}_{R}(v,u)$. Here we give the non-equilibrium scaling function
only in the large time separation limit $u\to \infty$:
\begin{eqnarray} \label{FR_inf}
\hspace*{0.cm}F^{\text{noneq}}_{R \infty}(v) &=& e^{-v} \left(
\int_{0}^{v} dt_2
\int_{0}^{t_2}  dt_1 {(e^{t_2-t_1}-1)} {\tilde\Sigma_{t_1 t_2}} +
\sigma \right)   \nonumber \\
{\tilde\Sigma_{t_1 t_2}} &=& \tau e^{\gamma_E} \frac{1}{(t_2 -
  t_1)^2} \left( 
\left(\frac{t_2 +
 t_1}{2 \sqrt{t_2 
    t_1}}\right)^{1-\epsilon} - 1 \right)
\end{eqnarray}
where the constant $\sigma = \int_1^{\infty} dt_2 \int_0^1 dt_1
{\tilde\Sigma_{t_1 t_2}}$ is a monotonic decreasing
function of $\epsilon$. The large $v$ behaviour is a
power law. The form (\ref{FR_inf}) is convergent due to the
substraction of the pole at $t_2=t_1$. The logarithmic
divergence associated to this pole yields a non trivial
$z$ exponent. The substracted piece precisely gives
the equilibrium scaling function for the response
$F_R^{\text{eq}}(v)$, up to ${\cal O}(\tau^2)$: 
\begin{eqnarray}\label{R_eq_app}
F_R^{\text{eq}}(v) = e^{- v} + \tau e^{\gamma_E} ((v-1) \Ei{(v)} e^{- v} + e^{- v} - 1)
\end{eqnarray}
where $\Ei{(v)}$ is the exponential integral. The
same result is also obtained if $t_i$ is taken to
be $t_i=-\infty$ from the outset (e.g. at large but
fixed system size), showing that
the nonequilibrium regime of the scaling function merges smoothly with
the equilibrium one \cite{foot4}. 

We now turn to correlation functions. To obtain the
equilibrium correlation one can simply use FDT 
which holds in this regime (i.e. ${\cal X}^q_{tt'} = 1$) and 
$\tilde{{\cal C}}^{q \text{eq}}_{tt'} = T
q^{-2} F_C^{\text{eq}}(v)$, with $\partial_v F_C^{\text{eq}}(v) = -
F_R^{\text{eq}}(v)$. One finds that at equilibrium and to
this order in $\tau$, $\tilde{{\cal C}}$ and ${\cal C}$
coincide. 

The non equilibrium connected correlation is
already non trivial   
in the absence of disorder \cite{leto_pure}. It
also takes a scaling form $\tilde C^q_{tt'} = {T}{q^{-2}} F_{\tilde
  C}^0(q^2 (t-t'),t/t')$ 
with $F_{\tilde C}^0(v,u)=e^{-|v|} - e^{-v \frac{u+1}{u-1}}$, $z=2$.
The FDR  
$X^q_{tt'} = (1 + e^{- 2\frac{v}{u-1}})^{-1}$ interpolates
between $1$ and $1/2$ as $u=t/t'$ increases from $1$ to
$\infty$ (for $q\neq 0$), while $X^{q=0}_{tt'}=\frac{1}{2}$
the ``random walk'' value. In presence of disorder one can solve
(\ref{expr_cc_dis_ci}) 
perturbatively in $\tau$ using the above solution for
the response. One obtains 
that $\tilde{\cal C}^q_{tt'}$, in the scaling regime, is consistent
with the scaling form \cite{footnote2}
\begin{equation}\label{scaling_cc}
\tilde{\cal C}^q_{tt'} = \frac{T}{q^2} \left(\frac{t}{t'}
\right)^{\theta_{\tilde{C}}} F_{\tilde{C}}(q^z (t-t'),t/t')
\end{equation}
The calculation
\cite{schehr_new} shows that
$\theta_{\tilde{C}} = \theta_{R}$, yielding  
an autocorrelation exponent $\lambda_{\tilde{C}} = - z (\theta_{\tilde
C}-1)$:
\begin{equation}\label{lambda_c_tilde}
\lambda_{\tilde{C}} =  2 + {\cal O}(\tau^2)
\end{equation} 
This value of $\theta_{\tilde{C}}$ is also such that $\lim_{u \to \infty} u
F_{\tilde{C}}(v,u) = F_{\tilde{C}\infty}(v)$.
We have obtained the complete expression of $F_{\tilde C}(v,u)$
\cite{schehr_new} 
from which we can extract the large $u$ behaviour. We
then discover the relation, valid for {\it any} $\epsilon$ 
\begin{eqnarray}\label{FCC_inf}
&&F_{\tilde{C}\infty}(v) = (2 + 4e^{\gamma_E}\tau) v F_{R\infty}(v) + {\cal
    O}(\tau^2) 
\end{eqnarray}

Having determined both response and correlation we
now obtain, in the scaling regime, the FDR (\ref{def_FDR}).
It is also characterized by the universal scaling function 
\begin{eqnarray}
\frac{1}{{\cal X}^q_{tt'}} = F_X(q^z(t-t'),t/t')
\end{eqnarray} 
which also depends on $\epsilon$ and
has a complicated form \cite{schehr_new}. In the
limit of large time separation i.e. $q^z(t-t')$
fixed, $t/t' \to \infty$, using
(\ref{FR_inf}, \ref{FCC_inf}) it simplifies into:
\begin{eqnarray}\label{FX_inf}
\lim_{u \to
  \infty}\frac{1}{{\cal X}^q_{tt'}}
 = 2 + 2 e^{\gamma_E} \tau + {\cal O}(\tau^2) = \frac{1}{X_\infty}
\end{eqnarray}  
independently of $v$ and $\epsilon$, i.e. of (small) wavevector 
and initial condition. There is thus a non trivial
asymptotic FDR in the CO glass phase, which similarly
to the case of pure critical points is in the interval $[1/2,0[$. 
Here however it continuously depends on temperature $T$. 

As in previous works \cite{calabrese_on, mayer_fdr_ising, randommass2}
one can also examine the ``diffusive'' 
mode $q = 0$. It is possible to obtain a simple
analytical form for any $u=t/t'$ in the case $\epsilon = 0$
(flat initial conditions $\phi(x,t=0)=0$), up to order ${\cal O}(\tau^2)$
\begin{equation}
\frac{1}{{\cal X}^{q=0}_{tt'}} = F^{\text{diff}}_{X}(u) = 2 + 2 \tau
e^{\gamma_E} \left(1 -
\atanh{\left(\frac{1}{\sqrt{u}}\right)}\right)
\end{equation}
Although this quantity depends on 
$\epsilon$ in general it reaches an $\epsilon$-independent
limit for $u \to \infty$. It is also interesting
to compute ${\cal X}^{x=0}_{tt'}$ in real space. 
In previous works \cite{calabrese_on}, it was argued that 
\begin{equation}
\lim_{t' \to \infty}\lim_{t \to \infty} {\cal X}^{x=0}_{tt'} =
\lim_{t' \to \infty} \lim_{t
\to \infty} {\cal X}^{q=0}_{tt'} = X_{\infty} 
\end{equation}
The $v$-independence found above (\ref{FX_inf})
puts in the present case  these heuristic arguments on
firmer grounds \cite{calabrese_on}. 

Following the discussion in
\cite{leto_leshouches}, this $X_{\infty}$ leads to an
effective temperature $T_{\text{eff}} = T/X_{\infty}$, that can be  
measured by a thermometer coupled to the field $\phi(x,t)$. 
Indeed, Fourier transforming
Eq. (\ref{scaling_resp}, \ref{scaling_cc}), 
one checks that the {\it local} ${\cal R}^{x=0}_{tt'}$ and ${\cal
\tilde{C}}^{x=0}_{tt'}$ are precisely of the form given in
\cite{footnote_scal_teff}.    

It turns out that it is crucial to consider the connected
correlation to obtain a non trivial FDR. We have also
performed the calculation \cite{schehr_co_pre} for the
correlation function ${\cal C}^q_{tt'}$ (\ref{def_C}). 
It exhibits a scaling form similar to (\ref{scaling_cc})
which decreases more slowly that $\tilde {\cal C}^q_{tt'}$
for large $u$. If we impose $\lim_{u \to \infty} u F_{C}(v,u) 
= F_{ C \infty}(v)$ one finds 
, for $\epsilon=0$
that $\theta_C = \theta_{\tilde{C}} + \frac{1}{2}$, leading to  
$\lambda_C = 1 - e^{\gamma_E}\tau + {\cal O}(\tau^2)$. 
The FDR is found to be 
${{\cal X}^{q=0}_{tt'}} = (2 + \tau e^{\gamma_E} \sqrt{u})^{-1}$ 
and thus does not seem to approach (uniformly in $\tau$) a non trivial
limit at large $u$. 
Thus, although both correlations give the same equilibrium
result to order ${\cal O}(\tau)$, only the connected one, as defined
here, yields 
a non trivial asymptotic FDR in the non equilibrium regime. For
$\epsilon \neq 0$, the large $u$ behavior is dominated by the 
initial condition  
and $F_C(v,u) \sim F_{C \infty}(v)$ \cite{picone_autoresponse, ci}.
The present considerations are also of interest for pure
systems when initial conditions are non zero, e.g. in a quench from an
ordered phase  
(or correlated initial conditions $\epsilon \neq 0$)
\cite{berthier_xy,ci,picone_autoresponse}. 
Indeed, in that case one can distinguish connected and non-connected
correlations (\ref{def_C}) which can also lead to different
behaviours of the FDR.

We have found a physically
relevant glass
phase in which one can show analytically the
existence of a non trivial FDR. It is a robust quantity 
independent of the initial condition under study and, for some
observables appear to be related to an effective temperature
$T_{\text{eff}}$.    
Its continuous dependence on $T$ reflects the marginal character of this
glass phase. Our analytical predictions 
can be tested in numerical 
simulations \cite{schehr_num}. They are in good agreement near
$T_g$ and it would be interesting to investigate
the aging behaviour at lower temperature, where
hints of some new physics already appear in the statics.

GS acknowledges the financial support provided
through the European Community's Human Potential Program 
under contract HPRN-CT-2002-00307, DYGLAGEMEM.


\begin{thebibliography}{32}
\bibitem{footnotenew}
It exhibits standard mean field phenomenology, with a non trivial
$X = T/T_g$ (and its static equivalent \cite{brG}) for each mode 
$q$, somewhat different from the picture emerging here from
the RG (where $q$ and $t$ must be scaled).
\bibitem{foot3}
all results here are thus valid 
for $\Lambda_l \ll q$, $\Lambda_l^z \ll 1/t'$
\bibitem{foot4}
since equilibrium is defined from
$t' \to \infty$ at fixed $\Lambda_l$, this is not
an obvious statement given \cite{foot3}.
\end{thebibliography}
\end{document}